\documentclass[twocolumn]{aastex631}
\usepackage{epsfig}
\renewcommand{\arraystretch}{1.2}
\received{Month DD, 2022}
\revised{Month DD, 2022}
\accepted{Month DD, 2022}
\submitjournal{ApJ}
\shorttitle{Prominence observations with ALMA}
\shortauthors{Heinzel et al.}

\begin{document}
\title{ALMA as a prominence thermometer: First observations}

\correspondingauthor{Petr Heinzel}
\email{pheinzel@asu.cas.cz}

\author[0000-0002-5778-2600]{Petr Heinzel}
\affiliation{Astronomical Institute, The Czech Academy of Sciences, 25165 Ond\v{r}ejov, Czech Republic}
\affiliation{University of Wroc{\l}aw, Center of Scientific Excellence - Solar and Stellar Activity, Kopernika 11, 51-622 Wroc{\l}aw, Poland}

\author[0000-0002-6505-4478]{Arkadiusz Berlicki} 
\affiliation{University of Wroc{\l}aw, Center of Scientific Excellence - Solar and Stellar Activity, Kopernika 11, 51-622 Wroc{\l}aw, Poland}
\affiliation{Astronomical Institute, University of Wroc{\l}aw, 51-622 Wroc{\l}aw, Poland}
\affiliation{Astronomical Institute, The Czech Academy of Sciences, 25165 Ond\v{r}ejov, Czech Republic}

\author[0000-0002-5092-9462]{Miroslav B\'{a}rta} 
\affiliation{Astronomical Institute, The Czech Academy of Sciences, 25165 Ond\v{r}ejov, Czech Republic}

\author[0000-0002-3898-1619]{Pawe{\l} Rudawy} 
\affiliation{Astronomical Institute, University of Wroc{\l}aw, 51-622 Wroc{\l}aw, Poland}

\author[0000-0003-3889-2609]{Stanislav Gun\'{a}r} 
\affiliation{Astronomical Institute, The Czech Academy of Sciences, 25165 Ond\v{r}ejov, Czech Republic}

 \author[0000-0002-4638-157X]{Nicolas Labrosse} 
 \affiliation{Department of Physics and Astronomy, University of Glasgow, Glasgow G12 8QQ, United Kingdom}

\author[0000-0003-0310-1598]{Krzysztof Radziszewski} 
\affiliation{Astronomical Institute, University of Wroc{\l}aw, 51-622 Wroc{\l}aw, Poland}

\begin{abstract}
We present first prominence observations obtained with ALMA 
in Band 3 at the wavelength of 3 mm. High-resolution
observations have been coaligned with the MSDP
H$\alpha$ data from Wroc{\l}aw-Bia{\l}k\'{o}w large coronagraph at
similar spatial resolution. We analyze one particular
co-temporal snapshot, first calibrating both ALMA and MSDP data and then
demonstrating a reasonable correlation between both. In particular
we can see quite similar fine-structure patterns in both ALMA brightness
temperature maps and MSDP maps of H$\alpha$ intensities.
Using ALMA we intend to derive the prominence kinetic temperatures. 
However, having current observations only in one band, we use an 
independent diagnostic
constraint which is the H$\alpha$ line integrated intensity.
We develop an inversion code and show that it can provide realistic
temperatures for brighter parts of the prominence where one gets
a unique solution, while within faint structures such inversion is
ill conditioned. In brighter parts ALMA serves as a prominence thermometer,
provided that the optical thickness in Band 3 is large enough. 
In order to find a relation between brightness and kinetic temperatures for
a given observed H$\alpha$ intensity, we constructed an extended grid of
non-LTE prominence models covering a broad range of prominence parameters.
We also show the effect of the plane-of-sky filling factor on our results.
\end{abstract}


\section{Introduction}

Spectral diagnostics of the prominence plasmas is being carried out for decades and is continuously improving 
thanks to new high-resolution ground and space instruments and more and more sophisticated non-LTE (i.e. departures from
Local Thermodynamic Equilibrium) modeling
techniques \citep{Labrosse+2010, VialEngvold2015}. However, based predominantly on analysis of spectral-line
intensities and shapes, it poses a difficult task to discriminate between various competing processes. Line intensity
depends on kinetic temperature, but also on density, prominence illumination and other factors and only a set
of suitably selected lines of different atomic species can provide  a well-constrained diagnostics. Thermal broadening
of lines is also competing with non-thermal ones, and namely with the so-called microturbulence. Several recent studies are based on multi-line  diagnostics, e.g. \citet{Park+2013}, \citet{Ruan+2019}, \citet{Okada+2020}, \citet{Peat+2021}, but the temperature is
always determined indirectly, sometimes using specific assumptions. However, we need to know the plasma kinetic
temperature as precisely as possible in order to discriminate between various heating and cooling processes 
which are still poorly understood \citep{Gilbert2015}. One promising way how to determine the temperature more directly is to
use the continuum observations and namely those in the radio domain where the Rayleigh-Jeans limit greatly
simplifies the analysis. This was already considered in the past (see a short review in \citealt{Heinzel+2015} and in \citealt{Wedem+2016}), but only recently the Atacama Large Millimeter Array (ALMA) started to provide radio observations with
much better spatial resolution, comparable with that in the optical domain. In order to predict the visibility
of prominences in various ALMA spectral bands, \citet{Heinzel+2015} used the H$\alpha$ coronagraphic observations
obtained at the Astronomical Observatory of the University of Wroc{\l}aw in Bialk\'{o}w, Poland and converted 
the optical images to ALMA ones. ALMA is often considered to be a plasma 'thermometer', however as shown
in \citet{Heinzel+2015} the prominences and their fine structures are in most cases only moderately optically thick in ALMA
radio bands and thus their brightness temperature is lower or much lower than the kinetic one. In order to
determine the optical thickness, one can use the integrated H$\alpha$ intensity which well correlates with
the plasma emission measure. Therefore developing a suitable technique one can convert the
brightness temperature into realistic kinetic one, using the integrated H$\alpha$ intensity to constrain the optical
thickness.  The obvious advantage of the continuum diagnostics is that it doesn't depend on plasma dynamics.
We may also assume that the {\em integrated} intensity of the H$\alpha$ line is only weakly sensitive to line broadening. Of course, that is not true for the shape of the H$\alpha$ line profiles.

Motivated by our previous study \citep{Heinzel+2015}, we have been proposing coordinated ALMA and H$\alpha$ 
observations during several ALMA cycles in frame of  different
teams. Finally, such observations took place on April 19, 2018 - see Section~\ref{sect:alma}. The advantage of these observations 
is that they are recorded almost simultaneously in radio and H$\alpha$ bands. These observations have been
partially analysed in PhD thesis of A. Rodger \citep{Rodger2019}. In the present
exploratory paper we performed complex ALMA data processing and calibration using an extended experience
of European ALMA Regional Center (ARC) at the Astronomical Institute of Czech Academy of Sciences in Ond\v{r}ejov. 
We then coaligned our H$\alpha$ observations with ALMA and developed an inversion technique to determine 
the prominence kinetic temperature. This technique is based on an extensive grid of non-LTE prominence models constructed specifically for this study.

\section{ALMA Observation}
\label{sect:alma}

We have analysed observations of the prominence performed on April 19, 2018,
in frame of the project 2017.0.01138.S (PI N. Labrosse). For the sake of direct
comparison with the H$\alpha$ data (see Section~\ref{sect:ha}), we have chosen just
a small subset of the collected data publicly available in the ALMA
archive. Namely, the execution blocks (EBs) 
\texttt{uid-A002-Xcc3ae3-Xb15e}
-- a small interferometric (INT) mosaic (5~pointings) in ALMA
\textit{Band~3} (continuum at $\approx$ 3mm, in the frequency ranges $93.0$ --- $97.0$~GHz and $103.0$---$107.0$~GHz),
and the complementary total-power (TP) scan of the entire solar disk --
\texttt{uid-A002-Xcc3ae3-Xb1f5} .
These EBs are co-temporal with the H$\alpha$
observations.

The INT data has been calibrated and imaged in CASA by the standard
procedure described in \citet{Shimojo+:2017} and the TP data according to
\citet{White+:2017}. For the imaging of the INT measurement set we have
reduced our selection to the shortest possible time interval $\approx$37.5~s
required for the full-mosaic scan between 15:38:07 -- 15:38:44~UT. The reason for this short time-scale imaging is rooted in the prominence dynamics that can blur the image: longer integrations exhibit visibly less fine structures.
The INT and TP data has been subsequently combined by CASA task
\texttt{feather()}
in order to recover low-frequency Fourier components
missing in the INT data. Beforehand, the necessary pre-requisities have been
performed: re-calculation of the TP image to Jy/beam, its shift to the same
reference frame as the INT one, cropping the relevant region in the TP data and
regrid, removal of the primary-beam (PB) correction in the INT image, etc.. We
have also cropped both (INT + TP) images to the same size (a sub-image of INT)
in order to keep just the data with high-enough signal (removing the border
where the PB gain is low). The \texttt{feather()} task 
takes care of proper handling of the overlapping range of Fourier components
of the TP and INT images. Eventually, the final combined image (brightness)
has been rescaled from Jy/beam to Kelvins and the reference frame transformed
form celestial RA/dec to the solar HPC as described in \citet{Skokic+Brajsa}.

INT/TP combination of the solar data has its caveats, especially for the limb
observations. The disparity between the two signals and from that implying
necessity to scale them properly has been reported in
\citet{Alisandrakis+:2017}. The \texttt{feather()} task allows to match the INT
and TP data by fitting two free parameters. We have used this option and
scaled the data
basically in line with the procedure of \citep{Alisandrakis+:2017}
in such a way that the combined image (1) has the average 
brightness temperature in the quiet solar region selected farther from the limb
corresponding to that obtained from the purely TP data, and (2) the average 
brightness temperature of the corona (i.e., the small off-limb region selected
farther from the prominence) is approximately zero (specifically, that it is not
significantly negative). 
For this sake we have tried several values of the \texttt{sdfactor} parameters in the CASA \texttt{feather()} task. The best fit of the above two requirements is reached for scaling factor of 0.8, which we have used in our combined image. Still somewhat slightly negative values of $T_{\rm b}$ are present in the corona, showing an imperfectness in the total flux reconstruction. However, significant distorsions (a rim of under- and over- shoots like in \citet{Shimojo+:2017}), are not visible.
Rotated and further cropped (with just the prominence
region selected) brightness-temperature map obtained by the above described
procedure is displayed in Fig. 1.
 
\section{MSDP H$\alpha$ Observations}
\label{sect:ha}

The same prominence was also observed in the optical range above the South-West part of the solar limb (PA=220$^{\circ}$) with the Large Coronagraph \citep{Gnevyshev1967, Rompolt1985} equipped with the Multi-channel Subtractive Double Pass (MSDP) imaging spectrograph \citep{Mein1991, Rompolt1994} at the Astronomical Observatory of the University of Wroc{\l}aw in Bialk\'{o}w, Poland. The observations of the prominence started at 10:02:41 UT, finished at 15:54:46 UT and were obtained in the scanning mode in the hydrogen H$\alpha$ line. In total, 739 scans were recorded.

The Large Coronagraph (LC) is a classical Lyot-type instrument with a 53 cm diameter main objective and nearly 14.5 m effective focal length. The MSDP imaging spectrographs provide two-dimensional images over the H$\alpha$ line profile \citep{Mein1977}. MSDP installed in Bialk\'{o}w has the  nine-channel prism-box. The prism-box forms nine images of the same area on the Sun (called channels), but shifted by $\Delta\lambda$ = 0.4 $\rm \AA$ in wavelength between the consecutive channels. The range of wavelengths band of a single channel is $\Delta\lambda$ = 4.3 $\rm \AA$. 

For each scan, after the reduction procedure, we constructed a set of twenty-three narrow-band nearly monochromatic images (waveband $\Delta\lambda$ = 0.06 $\rm \AA$) of the whole field-of-view (FOV) (up to $\pm$ 1.2$\rm \AA$ from the H$\alpha$ line centre), as well as the H$\alpha$ line profile at each pixel in the FOV. Spatial resolution of the obtained images is limited by seeing, on average to about 1.5 arcsec. The data was calibrated to the absolute units of the specific intensity using the quiet-Sun reference spectrum. In addition, to obtain correct values of the prominence absolute intensities in each pixel we removed the scattered light. That is particularly important for weak structures observed above the solar limb. Local intensities of the scattered light were evaluated for all observed altitudes above the solar limb, for all position angels within a segment of the ring with a height and width larger than the observed prominence, and for all analysed wavelengths using relevant data averaged over large structure-free regions located on both sides of the prominence. 

From the whole set of data we chose the scan observed from 15:37:56 UT, 11 seconds before time when the ALMA mosaic started to be reconstructed. The duration of the MSDP scan was 19 seconds. It is worth to notice that between 15:38:07 and 15:38:15 UT the prominence was observed really simultaneously with ALMA and MSDP. In the next step, for this spectral image, using calibrated H$\alpha$ line profiles for all pixels covering the prominence we computed the H$\alpha$ integrated intensities ($E$(H$\alpha$)) by integrating the line profiles within $\pm 1\rm \AA$ from the line centre. These intensities, together with ALMA brightness temperatures $T_{\rm b}$, are then used to determine the kinetic temperatures of the prominence plasma. The map of the prominence integrated intensities is shown in  Fig. 1.
%Fig.~\ref{fig:ALMA}

\begin{figure}
\centering
\includegraphics[width=0.45\textwidth]{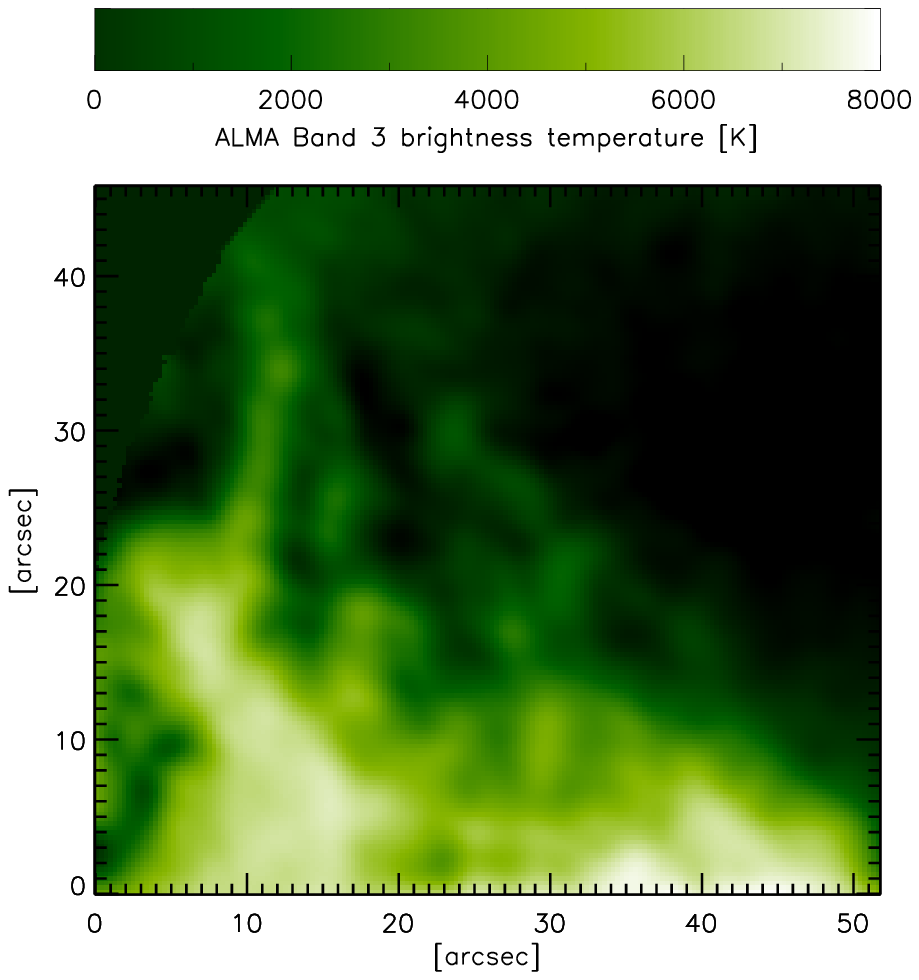}
\includegraphics[width=0.45\textwidth]{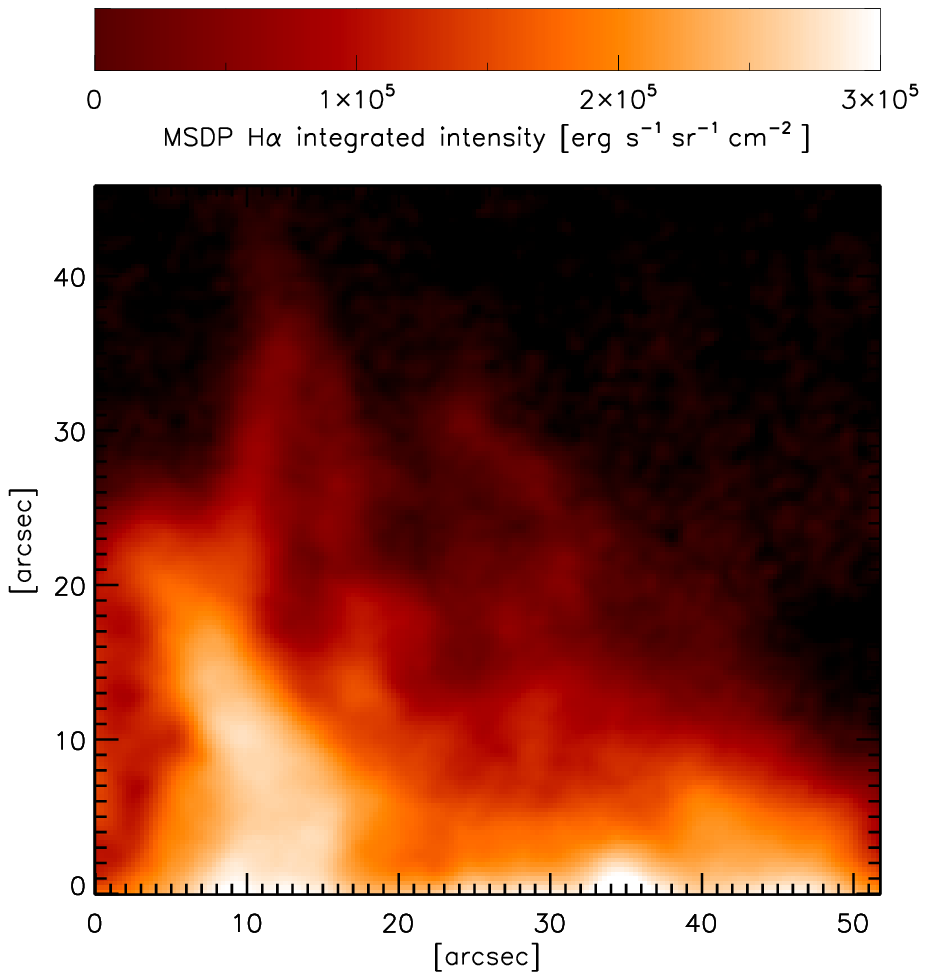}
\caption{Coaligned maps of the brightness temperature $T_{\rm b}$ obtained from calibrated 
ALMA Band 3 INT mosaic (top) and
H$\alpha$ integrated intensities from MSDP Wroc{\l}aw/Bialk\'{o}w (bottom). The 
spatial resolution of both images is comparable, between 1-2 arcsec. ALMA and MSDP data was 
recorded simultaneously, at 15:38:07 UT (start of ALMA mozaic) and 
15:37:56 UT (start of MSDP H$\alpha$ scan).
Individual fine structures are well recognised in both images.}
\label{fig:ALMA}
\end{figure}

\section{Formation of ALMA Continua in Prominences}
\label{sect:cont}

The radio continuum formation in prominences was reviewed in \citet{Heinzel+2015} where the reader can find all necessary
information.
The dominant source of opacity is the hydrogen thermal free-free continuum having the absorption coefficient in the form
\begin{equation}
\kappa_\nu(\mathrm{H})= 3.7\times 10^8 T^{-1/2} n_\mathrm{e} n_\mathrm{p} \nu^{-3} g_\mathrm{ff}   \, ,
\end{equation}
where $n_\mathrm{e}$ and $n_\mathrm{p}$ are the electron and proton densities, respectively,
$T$ is the kinetic temperature, $\nu$ is the frequency and $g_\mathrm{ff}$ is the Gaunt factor. 
In prominences H$^{-}$ free-free opacity can be neglected as estimated in \citet{Heinzel+2015}. 
Taking into account the correction for stimulated emission, one gets the expression often used
in the literature
\begin{equation}
\kappa_\nu =  0.018 n_\mathrm{e} n_\mathrm{p} T^{-3/2}\nu^{-2} g_\mathrm{ff} \, .
\end{equation}

The specific intensity $I_\nu$ for radiation normally emergent from the prominence on the limb, is expressed as
\begin{equation}
I_\nu = \int B_{\nu}(T) \mathrm{e}^{-t_\nu}\mathrm{d}t_\nu \, ,
\end{equation}
where $B_{\nu}(T)$ is the Planck source function, $t_{\nu}$ the optical depth. Although the continuum source
function is Planckian (for considered free-free processes),
the spectrum formation in not governed by LTE conditions.
Under characteristic prominence conditions, $n_{\rm e}$ and $n_{\rm p}$ depend on the non-LTE
radiative transfer mainly in hydrogen and helium continua where the photoionization
and radiative recombination processes play a dominant role for the
ionization equilibrium.

In the radio domain, $I_{\nu}$ and $B_{\nu}$ are directly proportional to the brightness temperature $T_{\rm b}$ and
to the plasma kinetic temperature $T_{\rm k}$, respectively
\begin{equation}
I_{\nu} = \frac{2 \nu^2 k}{c^2} T_{\rm b}  \quad B_{\nu} = \frac{2 \nu^2 k}{c^2} T _{\rm k} \, ,
\end{equation}
where $c$ is the speed of light.
Using this Rayleigh--Jeans law, the Equation (3) can be written as
\begin{equation}
T_{\rm b} = \int T_{\rm k} \mathrm{e}^{-t_\nu} {\rm d}t_{\nu} \, .
\end{equation}
Assuming a uniform kinetic temperature $T_{\rm k}$ along the LoS, we have for a prominence slab of
total optical thickness $\tau_\nu$
\begin{equation}
T_{\rm b} = f \, T_{\rm k} (1 -  \mathrm{e}^{-\tau_\nu}) \, ,
\end{equation}
where $\tau_\nu = \kappa_\nu \, L$ and $L$ is the {\it effective} geometrical thickness of the prominence.
$f$ is the plane-of-sky (PoS) filling factor to account for lower spatial resolution (e.g. \citealt{Irimajiri+1995}).
In the optically-thin limit we get simply $T_{\rm b} = f \, T_{\rm k} \tau(\nu)$.
At low spatial resolution we
always obtain a lower-limit estimate for $T_{\rm k}$ if the filling factor is neglected (i.e. set to unity), see also
our results below.

As we see from the above relations, the brightness temperature of the prominence at a given position and a given
frequency thus depends on the kinetic temperature, filling factor and optical thickness. The latter is directly proportional
to the emission measure EM=$n_{\rm e}^2 L$, provided that $n_{\rm e} = n_{\rm p}$ which applies to a pure hydrogen plasma.

\section{Prominence kinetic temperatures}
\label{sect:tkin}

\subsection{Forward modeling}
\label{sect:forw}
Our primary goal is to determine the prominence kinetic temperature independently
of the shapes of spectral lines which are often distorted by microturbulence and
other plasma dynamics. ALMA should provide ideal means for that because the radio continuum
emission is not affected by dynamics. Moreover, knowing the optical thickness
at a given frequency we can directly relate the brightness to kinetic temperatures
as described above. Observing the prominence simultaneously in at least two ALMA
bands (frequencies) where the continuum opacity substantially differs, one can get the 
kinetic temperature and optical thickness \citep[e.g.][]{Bastian+1993,Gunar+2018}.
However, in our present case 
we have only ALMA Band 3 coverage of the prominence and thus we have to use some additional
constraint. Assuming that the continuum optical thickness is proportional to total
emission measure along the line of sight EM, we can use a robust correlation between
such emission measure and integrated H$\alpha$ line intensity $E$(H$\alpha$). This correlation is
only weakly dependent on kinetic temperature as demonstrated by
\citet{Goutte+1993}. For a given measured $E$(H$\alpha$) we get slightly
different EM depending on the kinetic temperature. We use an extended grid of 1D-slab prominence models with fine steps of 100 K in kinetic temperatures (see below
for more detailed description of these models) to
search for the best fit between observed and synthetic $E$(H$\alpha$), for each
temperature separately. The step of 100 K is quite reasonable owing to calibration
inaccuracies of H$\alpha$. In this way we get EM for each temperature bin and from
that we can compute the continuum optical thickness and then the brightness temperature $T_{\rm b}$. 
We did this forward modeling for a few characteristic plasma temperatures and for the range of
$E$(H$\alpha$) as found in MSDP data. In Fig. 2a we plot all prominence pixels, together with
synthetic $T_{\rm b}$ computed under the assumption that the whole prominence has a
uniform kinetic temperature. This statistical analysis shows several interesting features. First,
we find a good correlation between ALMA $T_{\rm b}$ and MSDP $E$(H$\alpha$),
apart from some scatter, which indicates that both ALMA and MSDP observations were
reasonably well calibrated. Second, by plotting the synthetic curves of $T_{\rm b}$ versus $E$(H$\alpha$), we see that for $E$(H$\alpha$) $\ge$ 
10$^5$ ${\rm erg\;cm^{-2}\;s^{-1}\;sr^{-1}}$ (hereafter these units are referred as cgs) the brightness temperature
increases with increasing $T_{\rm kin}$ and we can see that the best fit of the latter is between 6000 and 7000 K.
On the other hand, for pixels with $E$(H$\alpha$) $\le$ 10$^5$ cgs the synthetic $T_{\rm b}$
is almost the same for all considered $T_{\rm kin}$. In this range the optical thickness
in Band 3 is small and thus the relation between $T_{\rm b}$ and $T_{\rm k}$
is rather ill conditioned meaning that we can not determine  the kinetic temperature with
a reasonable accuracy. The last interesting observation is that all pixels are located
below the color-coded curves which may look strange at the first sight. However, this indicates that
the location of the data points is not determined primarily by the data scatter. We interpret such a behaviour as due to
the ALMA plane-of-sky (PoS) filling factor $f$ lower than one. In Eq. 6 the r.h.s. is 
multiplied by $f$ and then the actual $T_{\rm b}$ is lower than in case of $f$=1 for
which the color-coded curves were synthesized. Interestingly, for $E$(H$\alpha$) $\le$ 10$^5$ cgs
one could estimate the value of $f$ just as the ratio of observed and color-coded $T_{\rm b}$,  simply
because the latter is almost insensitive to $T_{\rm kin}$ in the considered range. However, this
can not be done for $E$(H$\alpha$) $\ge$ 10$^5$ cgs where $T_{\rm b}$ splits according to $T_{\rm k}$,
the property which allows us to determine the kinetic temperature by inversions. In such case we have to assume
a certain value of $f$ as discussed below. Note that at largest  $E$(H$\alpha$) $T_{\rm b}$ saturates
to $T_{\rm k}$ as expected and under such conditions ALMA can be considered as a direct 
thermometer.

\begin{figure}
\centering
\includegraphics[width=0.5\textwidth]{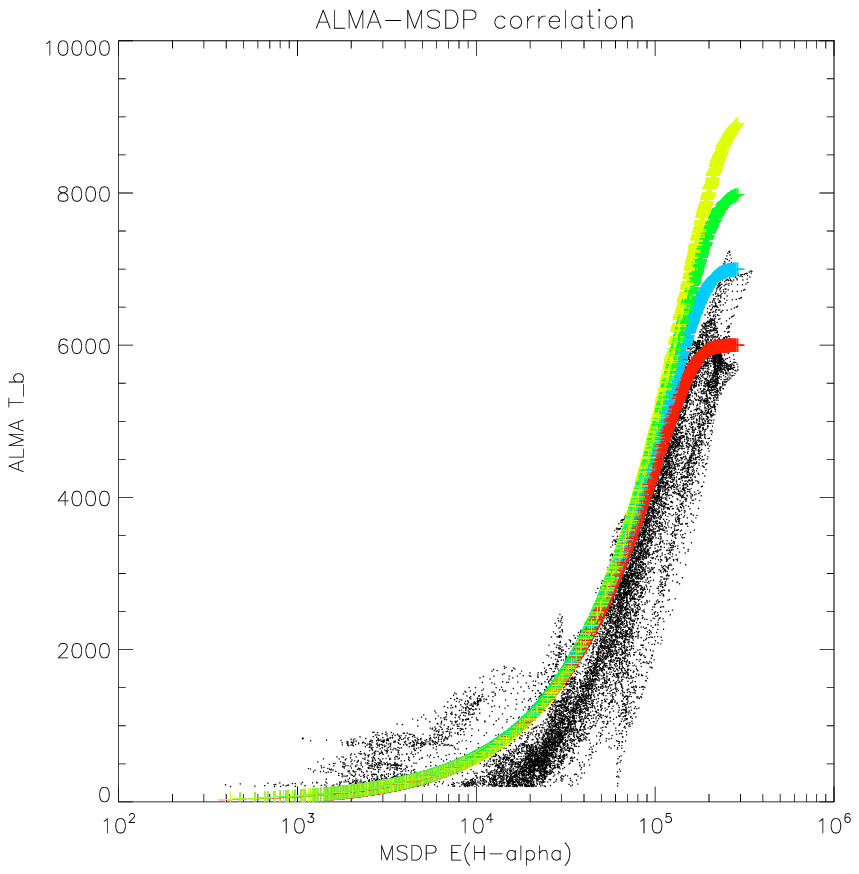}
\includegraphics[width=0.5\textwidth]{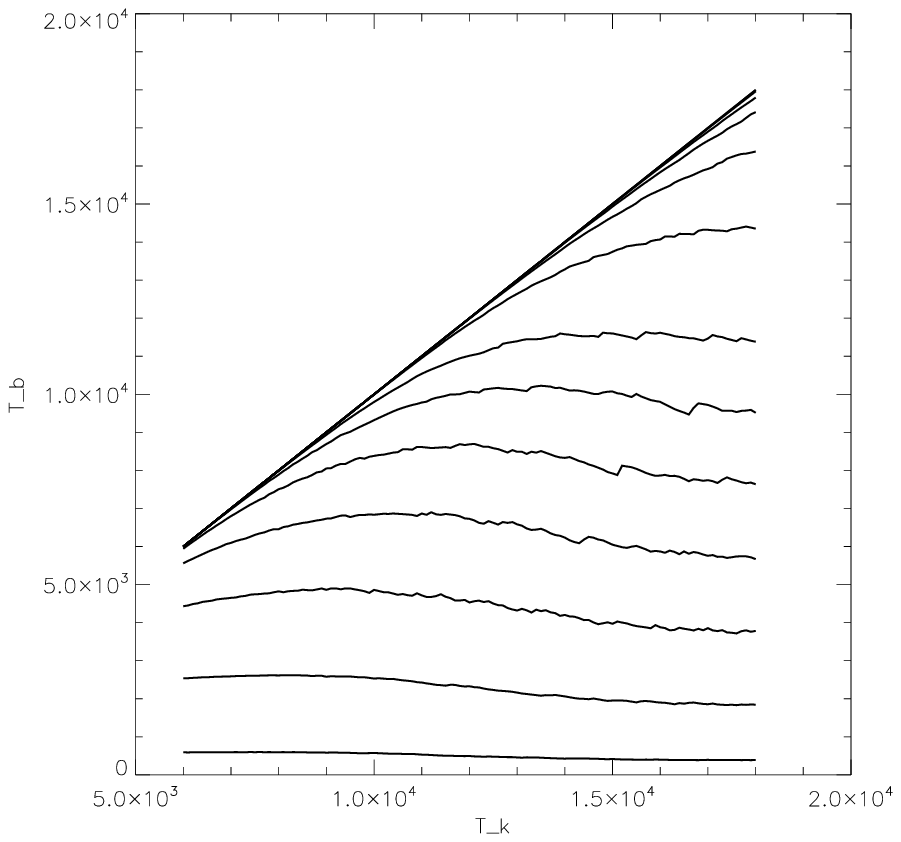}
\caption{a - relation between observed $T_{\rm b}$ and $E$(H$\alpha$) for all pixels from
Fig. 1 (dots). Color curves indicate the same relation constructed from our models and assuming a uniform
distribution of kinetic temperature in the whole FOV (yellow - 9000 K, green - 8000 K, blue - 7000 K, red - 6000 K).
Note that multiplying each color curve by certain value of $f$ one can see the effect of the filling factor. 
b - theoretical variations of Band 3 brightness temperature $T_{\rm b}$ versus  1D-slab 
kinetic temperature $T_{\rm k}$ for several values of the H$\alpha$ line integrated intensity. 
This plot was constructed from the grid of models (small kinks on some curves
are due to finite separation between the models in the grid).
The individual curves correspond (from below) to $E$(H$\alpha$) from 10$^4$ up to 10$^6$ ${\rm erg\;cm^{-2}\;s^{-1}\;sr^{-1}}$,
third curve from bottom is for our limiting value 10$^5$.}
\label{fig2} 
\end{figure}

\subsection{Inversions}
\label{sect:invers}
In \citet{Heinzel+2015} we proposed the temperature diagnostics using one selected ALMA band
and simultaneous H$\alpha$ observations. However, in the course of the present investigation
we realised that such method can work only for bright parts of prominences while for weaker parts
the inversion problem is ill-conditioned. 
This is demonstrated in Fig. 2b for several values of  $E$(H$\alpha$), using our grid of models. We see that for
$E$(H$\alpha$) $\le 10^5$ the curves are rather flat and then for a given observed $T_{\rm b}$ one obtains non-unique
solutions for  $T_{\rm k}$. Taking into account certain calibration inaccuracies, such inversion problem is
ill-conditioned.
We thus consider only the bright pixels with $E$(H$\alpha$) $\ge$ 10$^5$ cgs
and proceed as described above up to the point when we get variation of $T_{\rm b}$ versus $T_{\rm k}$
using the grid of models (i.e. variations like those shown in Fig. 2b). Then we search for the best fit between these  $T_{\rm b}$ and the
observed one divided by the filling factor $f$, at each pixel. Note that we apply the same filling factor also to MSDP observations which
have a similar spatial resolution as current ALMA Band 3. We thus divide the observed $E$(H$\alpha$) by the same value of $f$.
The resulting maps
of the plasma kinetic temperatures are shown in Fig. 3, for different values of $f$ which is taken to be
uniform for the whole FoV. For $f=1$ we have low observed $T_{\rm b}$ which causes that the inverted
$T_{\rm k}$ tend to reach lowest values in the grid (down to 5000 K). On the other hand, for $f=0.6$ the obtained $T_{\rm k}$ values may go to the upper limit of 12 000 K - we assume that the temperatures of cool parts of the
prominence do not exceed this value. Setting a higher limit would lead to non-unique inversions.
From the presented maps we conclude that a reasonable fit is obtained for $f$ around 0.8, but of course this is 
just an estimate because
different parts of the prominence may have different $f$ as seems to be the case from Fig. 2a.
The range of inverted temperatures for $f=0.8$ looks consistent with the behavior in Fig. 2a.
Note also that the inverted area somewhat expands when $f$
decreases which is due to extension of observed $E$(H$\alpha$) to lower values.

Low-temperature areas around the borders of the prominence are somewhat surprising, one would expect higher
temperature at the edges. However, there are three factors which may play a role. First, these areas are in fact not real
prominence edges because we limited our analysis only to brightest parts while in Fig. 1 we clearly see more extended
but fainter parts. Second, the pattern in our temperature maps is certainly influenced by assumed uniformity of the
filling factor $f$ - fainter structures may have much lower $f$ and thus their kinetic temperature will be in fact higher.
Finally, we limited our inversions by $T_{\rm k}$=12 000 K and relaxing this we could get two distinct solutions for
$T_{\rm k}$: one which tends toward the low temperatures (as in our maps) and the other which may have temperatures
larger than 12 000 K (see Fig. 2b). This kind of bifurcation must be constrained by other independent spectral diagnostics.

\begin{figure}
\centering
\includegraphics[width=0.38\textwidth]{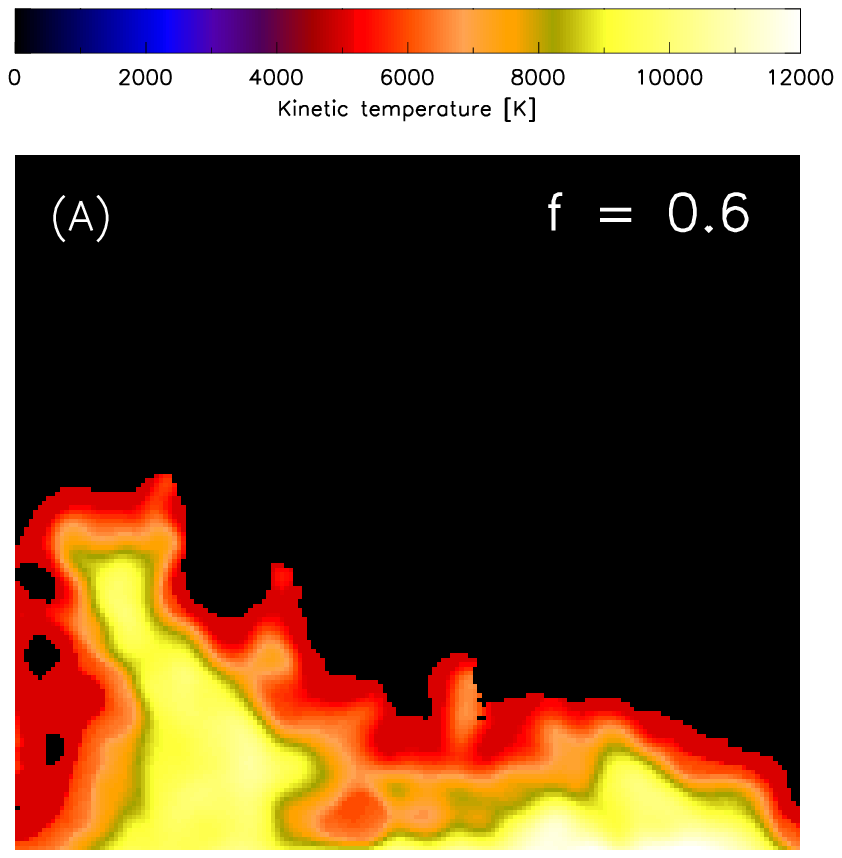}
\includegraphics[width=0.38\textwidth]{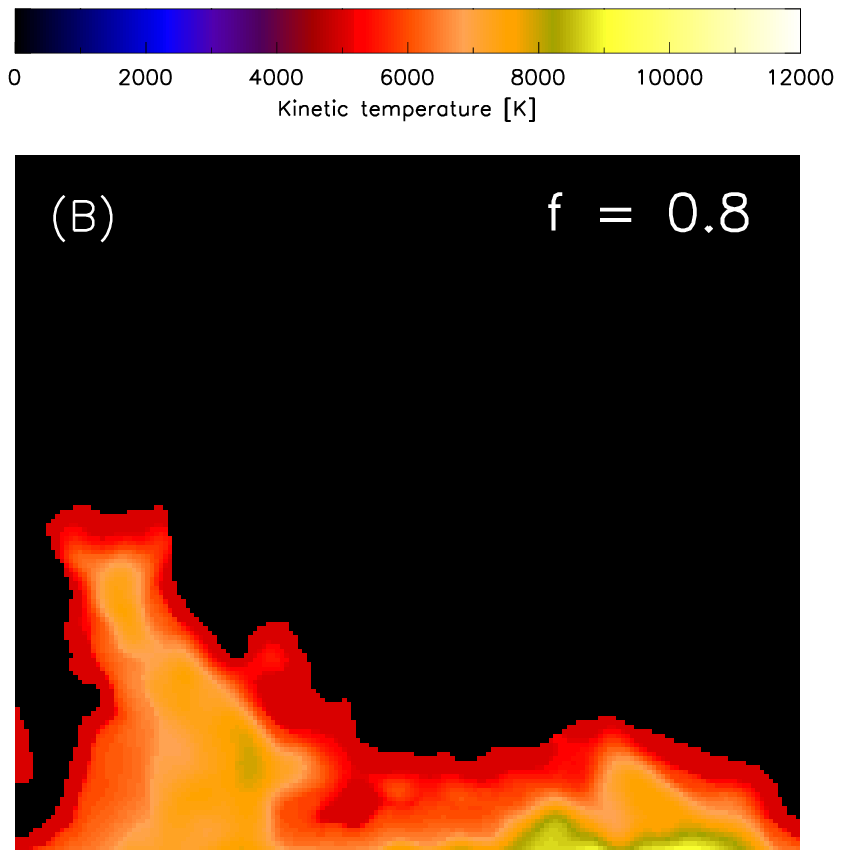}
\includegraphics[width=0.38\textwidth]{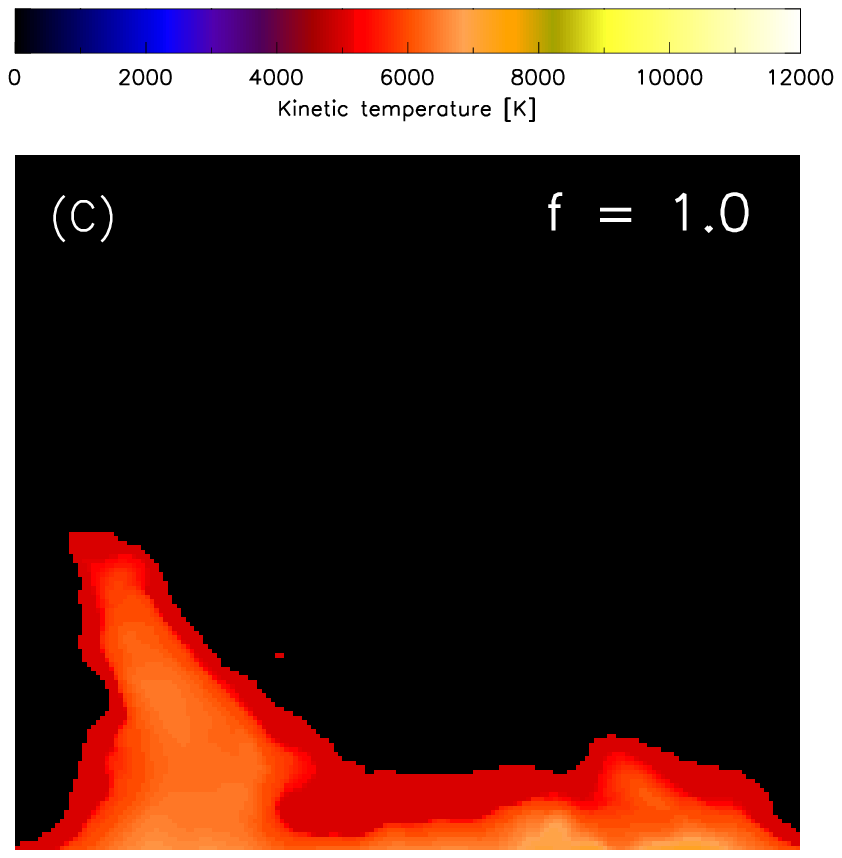}
\caption{Inverted maps of kinetic temperatures for different values of the filling factor:
A) $f$=0.6, B) $f$=0.8, C) $f$=1.0. The field-of-view is the same as in Fig. 1. A more extended prominence area at lower $f$
is due to weaker structures in H$\alpha$ taken into account - see the text.}
\label{fig3} 
\end{figure}

\subsection{Grid of non-LTE models}
\label{sect:grid}
Our extended grid of 1D-slab non-LTE models constructed for this work consists of 131 temperatures covering the range
from 5000 to 18000 K with the step of 100 K, it covers 
a reasonable range of gas pressures from 0.01 to 1.0 dyn cm$^{-2}$ and of
effective prominence thicknesses along the LoS from 500 to 5000 km with the step of 100 km. 
For each temperature we have 828 models and altogether we computed 131 $\times$ 828 = 108468
models. The non-LTE models are based on the MALI technique as described in \citet{Heinzel1995}, but here we used
the partial-redistribution for hydrogen Lyman-$\alpha$ and Lyman-$\beta$ lines. 
We use only one characteristic microturbulent velocity 5 km sec$^{-1}$ and one
mean height above the limb to specify the diluted incident radiation illuminating
the prominence slab on both sides. Here it is worth mentioning that although the ALMA continuum
source function is Planckian (Section 4), the emission measure, or more precisely the
product of electron and proton density times the effective thickness, results from the
non-LTE modeling of hydrogen ionization which represents very complex task undertaken here.
In this work we set $n_{\rm e}=n_{\rm p}$ because our radiative-transfer non-LTE code doesn't account for
ionization of other species and namely of the helium. This is quite reasonable in cool prominence regions where helium
doesn't contribute to the electron density more than typically 10 \%.

Our 1D-slab models are static in this exploratory work, while the real H$\alpha$
line profiles can be affected by plasma dynamics. Hydrogen lines are less sensitive
to microturbulent broadening than those of other heavier species but the line profiles
are still shaped by macroscopic fine-structure motions if present, as shown e.g. in \citet{Gunar+2012}.
However, we assume here that observed {\em integrated} line emission is not much influenced
by motions and thus that static models are adequate for this exploratory study. 

\section{Discussion and Future Prospects}

The principal aim of this exploratory work was to demonstrate the ability of ALMA to
provide diagnostics of the prominence kinetic temperature. Using observations currently
available only in a single band at a time, we demonstrate how the temperature can be derived based on additional constraint
provided by the co-temporal and co-spatial H$\alpha$ observations. The method described here works reasonably well for
bright prominence structures which become optically-thick in H$\alpha$ and also in the considered ALMA
band, but for weaker features the inversion is ill-conditioned, which is even worse if the data
uncertainties are considered. However, we found a large sensitivity of results on the adopted values of the PoS
filling factor, the issue to be further investigated. In fact even for weak structures the solution could be
obtained provided that they are weak due to very small filling factor. In our analysis we considered
a uniform filling factor over the whole FoV and have shown the inversions for its different values.
But in reality weak features may be due to very small factor, while brightest ones may have filling
factor closer to unity. For future ALMA observations
it would be useful to get data in several (at least two rather distinct) bands and test
such multi-band diagnostics against the one developed here. We are fully
aware of simplifications in our non-LTE modeling used for this work and plan to extend
it to a fine-structure case including the Prominence-Corona Transition Region (PCTR) as previously done for other diagnostics. But as
discussed above, the fine-structure dynamics is not affecting our results based on continuum
observations, only some marginal influence can be expected on integrated H$\alpha$ emission.
Fully 3D modeling of ALMA visibility of heterogeneous prominences was already performed in
\citet{Gunar+2016} and \citet{Gunar+2018}, but the next step is to consider relevant inversion strategies. The principal conclusion
here is that ALMA provides reliable data for prominence temperature diagnostics and these data
give the results which are consistent with the H$\alpha$ observations. Our diagnostics here was 
shown to be limited to certain range of H$\alpha$ intensities,
otherwise the inversions are problematic. Possible solution is to add another independent diagnostics based
on lines of other species like CaII and MgII (see e.g. chromospheric inversions of 
\citet{Santos+2018}). The ALMA community is continuously
developing methods of data reduction and calibration and specifically for off-limb targets this
represents a challenge because of the presence of sharp limb. Contrary to previous expectations, the current
ALMA Band 3 prominence observations have an order of magnitude lower resolution which is 
comparable to H$\alpha$ MSDP observations or somewhat better (see Fig. 1). Future prominence observations
should be optimised for higher spatial resolution and (quasi)-simultaneously in at least two distinct bands. 

\begin{acknowledgements}
The authors are indebted to the anonymous referee for his/her constructive comments.
 P.H. and S.G. acknowledge support from the grants GACR 19-16890S and 19-17102S of the 
Czech Science Foundation. P.H., M.B. and S.G. thank for the support from project 
RVO:67985815 of the Astronomical Institute of the Czech Academy of Sciences. P.H. and A.B. 
were supported by the program "Excellence Initiative - Research University" for years 
2020-2026 at University of Wroc{\l}aw, project no. BPIDUB.4610.96.2021.KG.
The work of A.B. was partially supported by the program "Excellence Initiative - Research 
University" for years 2020-2026 at University of Wroc{\l}aw, project no. BPIDUB.4610.15.2021.KP.B.
This paper makes use of the following ALMA data: ADS/JAO.ALMA\#2017.0.01138.S. We appreciate
the effort of PI's team to facilitate these ALMA observations.
ALMA is a partnership of ESO (representing its member states), NSF (USA) and
NINS (Japan), together with NRC (Canada), MOST and ASIAA (Taiwan), and KASI
(Republic of Korea), in cooperation with the Republic of Chile. The Joint ALMA
Observatory is operated by ESO, AUI/NRAO and NAOJ. M.B. acknowledges support from the grant 
GACR 20-09922J of the Czech Science
Foundation. ALMA data reduction has been facilitated by the Czech node of the
EU ARC, supported by the Czech Ministry of Education via the project No. LM2015067.
\end{acknowledgements}

\facility{ALMA, Large Coronagraph/MSDP of the Astronomical Observatory of the University of Wroc{\l}aw}

%\bibliographystyle{aasjournal}
%\bibliography{heinzel_alma.bib}

\begin{thebibliography}{}
\expandafter\ifx\csname natexlab\endcsname\relax\def\natexlab#1{#1}\fi
\providecommand{\url}[1]{\href{#1}{#1}}
\providecommand{\dodoi}[1]{doi:~\href{http://doi.org/#1}{\nolinkurl{#1}}}
\providecommand{\doeprint}[1]{\href{http://ascl.net/#1}{\nolinkurl{http://ascl.net/#1}}}
\providecommand{\doarXiv}[1]{\href{https://arxiv.org/abs/#1}{\nolinkurl{https://arxiv.org/abs/#1}}}

\bibitem[{{Alissandrakis} {et~al.}(2017){Alissandrakis}, {Patsourakos},
  {Nindos}, \& {Bastian}}]{Alisandrakis+:2017}
{Alissandrakis}, C.~E., {Patsourakos}, S., {Nindos}, A., \& {Bastian}, T.~S.
  2017, \aap, 605, A78, \dodoi{10.1051/0004-6361/201730953}

\bibitem[{{Bastian} {et~al.}(1993){Bastian}, {Ewell}, \&
  {Zirin}}]{Bastian+1993}
{Bastian}, T.~S., {Ewell}, M.~W., J., \& {Zirin}, H. 1993, \apj, 418, 510,
  \dodoi{10.1086/173413}

\bibitem[{{da Silva Santos} {et~al.}(2018){da Silva Santos}, {de la Cruz
  Rodr{\'\i}guez}, \& {Leenaarts}}]{Santos+2018}
{da Silva Santos}, J.~M., {de la Cruz Rodr{\'\i}guez}, J., \& {Leenaarts}, J.
  2018, \aap, 620, A124, \dodoi{10.1051/0004-6361/201833664}

\bibitem[{{Gilbert}(2015)}]{Gilbert2015}
{Gilbert}, H. 2015, in Astrophysics and Space Science Library, Vol. 415, Solar
  Prominences, ed. J.-C. {Vial} \& O.~{Engvold}, 157,
  \dodoi{10.1007/978-3-319-10416-4\_7}

\bibitem[{{Gnevyshev} {et~al.}(1967){Gnevyshev}, {Nikolsky}, \&
  {Sazanov}}]{Gnevyshev1967}
{Gnevyshev}, M.~N., {Nikolsky}, G.~M., \& {Sazanov}, A.~A. 1967, \solphys, 2,
  223, \dodoi{10.1007/BF00155923}

\bibitem[{{Gouttebroze} {et~al.}(1993){Gouttebroze}, {Heinzel}, \&
  {Vial}}]{Goutte+1993}
{Gouttebroze}, P., {Heinzel}, P., \& {Vial}, J.~C. 1993, \aaps, 99, 513

\bibitem[{{Gun{\'a}r} {et~al.}(2018){Gun{\'a}r}, {Heinzel}, {Anzer}, \&
  {Mackay}}]{Gunar+2018}
{Gun{\'a}r}, S., {Heinzel}, P., {Anzer}, U., \& {Mackay}, D.~H. 2018, \apj,
  853, 21, \dodoi{10.3847/1538-4357/aaa001}

\bibitem[{{Gun{\'a}r} {et~al.}(2016){Gun{\'a}r}, {Heinzel}, {Mackay}, \&
  {Anzer}}]{Gunar+2016}
{Gun{\'a}r}, S., {Heinzel}, P., {Mackay}, D.~H., \& {Anzer}, U. 2016, \apj,
  833, 141, \dodoi{10.3847/1538-4357/833/2/141}

\bibitem[{{Gun{\'a}r} {et~al.}(2012){Gun{\'a}r}, {Mein}, {Schmieder},
  {Heinzel}, \& {Mein}}]{Gunar+2012}
{Gun{\'a}r}, S., {Mein}, P., {Schmieder}, B., {Heinzel}, P., \& {Mein}, N.
  2012, \aap, 543, A93, \dodoi{10.1051/0004-6361/201218940}

\bibitem[{{Heinzel}(1995)}]{Heinzel1995}
{Heinzel}, P. 1995, \aap, 299, 563

\bibitem[{{Heinzel} {et~al.}(2015){Heinzel}, {Berlicki}, {B{\'a}rta},
  {Karlick{\'y}}, \& {Rudawy}}]{Heinzel+2015}
{Heinzel}, P., {Berlicki}, A., {B{\'a}rta}, M., {Karlick{\'y}}, M., \&
  {Rudawy}, P. 2015, \solphys, 290, 1981, \dodoi{10.1007/s11207-015-0719-7}

\bibitem[{{Irimajiri} {et~al.}(1995){Irimajiri}, {Takano}, {Nakajima},
  {Shibasaki}, {Hanaoka}, \& {Ichimoto}}]{Irimajiri+1995}
{Irimajiri}, Y., {Takano}, T., {Nakajima}, H., {et~al.} 1995, \solphys, 156,
  363, \dodoi{10.1007/BF00670232}

\bibitem[{{Labrosse} {et~al.}(2010){Labrosse}, {Heinzel}, {Vial}, {Kucera},
  {Parenti}, {Gun{\'a}r}, {Schmieder}, \& {Kilper}}]{Labrosse+2010}
{Labrosse}, N., {Heinzel}, P., {Vial}, J.~C., {et~al.} 2010, \ssr, 151, 243,
  \dodoi{10.1007/s11214-010-9630-6}

\bibitem[{{Mein}(1977)}]{Mein1977}
{Mein}, P. 1977, \solphys, 54, 45, \dodoi{10.1007/BF00146423}

\bibitem[{{Mein}(1991)}]{Mein1991}
---. 1991, \aap, 248, 669

\bibitem[{{Okada} {et~al.}(2020){Okada}, {Ichimoto}, {Machida}, {Tokuda},
  {Huang}, \& {UeNo}}]{Okada+2020}
{Okada}, S., {Ichimoto}, K., {Machida}, A., {et~al.} 2020, \pasj, 72, 71,
  \dodoi{10.1093/pasj/psaa014}

\bibitem[{{Park} {et~al.}(2013){Park}, {Chae}, {Song}, {Maurya}, {Yang},
  {Park}, {Jang}, {Nah}, {Cho}, {Kim}, {Ahn}, {Cao}, \& {Goode}}]{Park+2013}
{Park}, H., {Chae}, J., {Song}, D., {et~al.} 2013, \solphys, 288, 105,
  \dodoi{10.1007/s11207-013-0271-2}

\bibitem[{{Peat} {et~al.}(2021){Peat}, {Labrosse}, {Schmieder}, \&
  {Barczynski}}]{Peat+2021}
{Peat}, A.~W., {Labrosse}, N., {Schmieder}, B., \& {Barczynski}, K. 2021, \aap,
  653, A5, \dodoi{10.1051/0004-6361/202140907}

\bibitem[{{Rodger}(2019)}]{Rodger2019}
{Rodger}, A. 2019, PhD thesis, University of Glasgow, United Kingdom

\bibitem[{{Rompolt}(1994)}]{Rompolt1994}
{Rompolt}, B. 1994, JOSO Annual Report 1993, 87

\bibitem[{{Rompolt} \& {Rudawy}(1985)}]{Rompolt1985}
{Rompolt}, B., \& {Rudawy}, P. 1985, Postepy Astronomii Krakow, 33, 141

\bibitem[{{Ruan} {et~al.}(2019){Ruan}, {Jej{\v{c}}i{\v{c}}}, {Schmieder},
  {Mein}, {Mein}, {Heinzel}, {Gun{\'a}r}, \& {Chen}}]{Ruan+2019}
{Ruan}, G., {Jej{\v{c}}i{\v{c}}}, S., {Schmieder}, B., {et~al.} 2019, \apj,
  886, 134, \dodoi{10.3847/1538-4357/ab4b50}

\bibitem[{{Shimojo} {et~al.}(2017){Shimojo}, {Bastian}, {Hales}, {White},
  {Iwai}, {Hills}, {Hirota}, {Phillips}, {Sawada}, {Yagoubov}, {Siringo},
  {Asayama}, {Sugimoto}, {Braj{\v{s}}a}, {Skoki{\'c}}, {B{\'a}rta}, {Kim}, {de
  Gregorio-Monsalvo}, {Corder}, {Hudson}, {Wedemeyer}, {Gary}, {De Pontieu},
  {Loukitcheva}, {Fleishman}, {Chen}, {Kobelski}, \& {Yan}}]{Shimojo+:2017}
{Shimojo}, M., {Bastian}, T.~S., {Hales}, A.~S., {et~al.} 2017, \solphys, 292,
  87, \dodoi{10.1007/s11207-017-1095-2}

\bibitem[{{Skoki{\'c}} \& {Braj{\v{s}}a}(2019)}]{Skokic+Brajsa}
{Skoki{\'c}}, I., \& {Braj{\v{s}}a}, R. 2019, arXiv e-prints, arXiv:1904.08263.
\newblock \doarXiv{1904.08263}

\bibitem[{{Vial} \& {Engvold}(2015)}]{VialEngvold2015}
{Vial}, J.-C., \& {Engvold}, O., eds. 2015, Astrophysics and Space Science
  Library, Vol. 415, {Solar Prominences}, \dodoi{10.1007/978-3-319-10416-4\_5}

\bibitem[{{Wedemeyer} {et~al.}(2016){Wedemeyer}, {Bastian}, {Braj{\v{s}}a},
  {Hudson}, {Fleishman}, {Loukitcheva}, {Fleck}, {Kontar}, {De Pontieu},
  {Yagoubov}, {Tiwari}, {Soler}, {Black}, {Antolin}, {Scullion}, {Gun{\'a}r},
  {Labrosse}, {Ludwig}, {Benz}, {White}, {Hauschildt}, {Doyle}, {Nakariakov},
  {Ayres}, {Heinzel}, {Karlicky}, {Van Doorsselaere}, {Gary}, {Alissandrakis},
  {Nindos}, {Solanki}, {Rouppe van der Voort}, {Shimojo}, {Kato},
  {Zaqarashvili}, {Perez}, {Selhorst}, \& {Barta}}]{Wedem+2016}
{Wedemeyer}, S., {Bastian}, T., {Braj{\v{s}}a}, R., {et~al.} 2016, \ssr, 200,
  1, \dodoi{10.1007/s11214-015-0229-9}

\bibitem[{{White} {et~al.}(2017){White}, {Iwai}, {Phillips}, {Hills}, {Hirota},
  {Yagoubov}, {Siringo}, {Shimojo}, {Bastian}, {Hales}, {Sawada}, {Asayama},
  {Sugimoto}, {Marson}, {Kawasaki}, {Muller}, {Nakazato}, {Sugimoto},
  {Braj{\v{s}}a}, {Skoki{\'c}}, {B{\'a}rta}, {Kim}, {Remijan}, {de Gregorio},
  {Corder}, {Hudson}, {Loukitcheva}, {Chen}, {De Pontieu}, {Fleishmann},
  {Gary}, {Kobelski}, {Wedemeyer}, \& {Yan}}]{White+:2017}
{White}, S.~M., {Iwai}, K., {Phillips}, N.~M., {et~al.} 2017, \solphys, 292,
  88, \dodoi{10.1007/s11207-017-1123-2}

\end{thebibliography}

\end{document}